# Mass enhancement in 3*d* and *s-p* perovskites from symmetry breaking

Zhi Wang, Oleksandr I. Malyi, Xingang Zhao, and Alex Zunger

Renewable and Sustainable Energy Institute, University of Colorado, Boulder, Colorado 80309

## Abstract

In some *d*-electron oxides the measured effective mass $m^*_{\text{exptl}}$ has long been known to be significantly larger than the model effective mass $m^*_{\text{model}}$ deduced from mean-field band theory, i.e., $m^*_{\text{exptl}} = \beta m^*_{\text{model}}$, where $\beta > 1$ is the "mass enhancement", or "mass renormalization" factor. Previous applications of density functional theory (DFT), based on the smallest number of possible magnetic, orbital, and structural degrees of freedom, missed such mass enhancement, a fact that was taken as evidence of strong electronic correlation being the exclusive enabling physics. The current paper reports that known modalities of energy-lowering symmetry-broken spin and structural effects included in mean-field DFT show mass enhancement for both electrons and holes in a range of *d*-electron perovskites SrVO₃, SrTiO₃, BaTiO₃, and LaMnO₃ as well as *p*-electron perovskites CsPbI₃ and SrBiO₃, including metals (SrVO₃) and insulators (the rest). This is revealed *only* when enlarged unit cells of the same parent global symmetry, which are large enough to allow for symmetry breaking distortions and concomitant variations in spin order, are explored for their ability to lower the total energy. The paper analyzes the contributions of different symmetry-broken modalities to mass enhancement, finds common effects in the range of *d*- as well as *p*-electron perovskites. Thus, symmetry breaking in mean-field theory could describe effects that were previously attributed exclusively to complex correlated treatments. Indeed, broken symmetries are often experimentally observed, *e.g.*, octahedral tilting, Jahn-Teller distortions, or bond disproportionation, and often provide intuitive explanations in terms of degeneracy removal due to lowering of structural symmetry, or as shifting of band energies in explainable directions.

# I. Introduction

The effective mass $m^*$ defined [1] as the reciprocal of the wave vector curvature $\partial^2 E/\hbar^2 \partial k_i \partial k_j$ of the band dispersion relation $E(k)$ (where $k_i$ and $k_j$ are wave vectors) is a central quantity in condensed matter physics, widely used to characterize the band structure, carrier transport, and wave function localization. Recently, this quantity has attracted attention in the context of *d*-electron correlated oxide physics, where the measured effective mass $m^*_{\text{exptl}}$ has been noted in some cases to be significantly larger than the model effective mass $m^*_{\text{model}}$ deduced from simplified mean-field band theory, $m^*_{\text{exptl}} = \beta m^*_{\text{model}}$, where $\beta$ is the 'mass enhancement', or 'mass renormalization' factor. Effective masses $m^*_{\text{exptl}}$ are generally deduced from experiment via model assumptions (such as band parabolicity or various averages over the mass tensor), leading to different effective mass definitions in different experiments, including the mass $m^* \propto 1/v_F$, deduced from Fermi-velocity ($v_F$), or from density of states (DOS) $m^* \propto (D(E))^{2/3}$, from specific-heat coefficient $m^* \propto \gamma$, from magnetic susceptibility $\chi \propto m^*\left(1 - \frac{m_0^2}{3m^{*2}}\right)$, and from bandwidth $W \propto 1/m^*$. Values of $\beta > 1$ were reported in the literature for Fe-based superconductors [2,3], halide perovskites [4], titanites [5–7], ruthenates [8–10], and vanadates [11–14], *etc*. Such mass enhancement factors deduced from experiment $\beta(\text{exptl/model}) = m^*_{\text{exptl}}/m^*_{\text{model}}$ with reference to a model calculation $m^*_{\text{model}}$ were compared with theoretical values $\beta(\text{Theory/model}) = m^*_{\text{Theory}}/m^*_{\text{model}}$ obtained from advanced theory (such as dynamic mean-field theory, DMFT [15–24]). Because $m^*_{\text{model}}$ is obtained from mean-field band theory, the predicted theoretical enhancement $\beta(\text{Theory/model}) > 1$ has been interpreted to be due to strong correlation effects [15–24]. For example, in DMFT, wavefunction localizes and bandwidth narrows (thus leading to mass enhancement) due to pure *electronic* symmetry breaking [24,25] induced by the dynamic self-energy from the impurity atom embedded in a mean-field bath. Finding for a compound that $\beta(\text{DMFT/model}) > 1$ consistent with $\beta(\text{exptl/model}) > 1$ helped classify the pertinent compounds as being highly correlated, given that mean-field theory has been argued unable to describe the mass enhancement.

This line of thinking, however, does not consider the possibility that mass enhancement could be described by methods other than those symmetry-restricted structures. Indeed, the model calculations used to extract $m^*_{\text{model}}$ have invariably been [15–24] rather naïve (N) level of density functional theory (N-DFT), based on the least number of possible magnetic, orbital, and structural degrees of freedom. Such calculations have assumed one or a few of the following approximations: A highly symmetric unit cell symmetry (*e.g.*, $Pm\bar{3}m$ cubic); a nonmagnetic (NM) spin configuration and no atomic displacements relative to the averaged high symmetry structure. The shortcomings of such simplistic N–DFT approaches are evident, among others in (i) predicting *metallic states* for known insulators (as illustrated for *e.g.*, binary 3*d* NiO, MnO, CoO, FeO insulators [26], ternary 3*d* oxide perovskites [27] LaVO$_3$, LaMnO$_3$, and YNiO$_3$, as well as other compounds [28,29] such as CaIrO$_3$, LaTiO$_3$, SrBiO$_3$, TiO$_{2-x}$, CeO$_{2-x}$, CuBi$_2$O$_4$, Sr$_2$IrO$_4$, Li$_x$TiO$_2$, and Ba$_4$As$_3$, and by the fact that (ii) they predict a total energy far higher (by ~1 eV per formula unit, eV/f.u.) than what conventional DFT [27] provides, as well as by (iii) missing the orbital order [27].

However, these results, obtained from rather a highly restricted version of DFT, do not necessarily reflect what DFT can do. The failure of these naïve applications of mean-field band theory to predict mass enhancement often goes hand in hand with such assumptions of simplistic, highly-symmetric minimal unit cell model containing only the least number of possible magnetic, orbital and structural degrees of freedom. For example, the cubic crystal structure of *halide perovskites* ABX$_3$ (A = Cs, MA, FA; B = Sn, Pb; X = Cl, Br, I) is described in X-ray diffraction databases as $Pm\bar{3}m$ cubic, having a single ABX$_3$ f.u. per cell, with all octahedra being ideally shaped, untitled, and oriented parallel to each other, representing a single BX$_6$ local motif (monomorphous structure). The $Pm\bar{3}m$ cubic structure has been extensively used [30–33] to calculate the standard electronic band structure as well as phonon lattice dynamics. However, in reality, the cubic phase of such ABX$_3$ perovskites often manifests *static* atomic distortions off Wyckoff positions as an intrinsic expression of their chemical bonding, as seen by local probes [34,35] and reproduced by static total energy optimization [36] even before the contribution of thermal motion sets in. Such symmetry breakings are often observables by X-ray local probe measurements such as the extended X-ray-absorption fine-structure (EXAFS) and pair distribution function (PDF) [37–42]. Such distortions are easy to miss in total energy minimization if one uses the *average* X-ray diffraction (XRD) $Pm\bar{3}m$ cubic cell of single f.u., because that such single-f.u. cell geometrically excludes the creation of a periodic lattice with tilted octahedra. Permitting a larger cell ("supercell") yet with constrained cubic lattice vectors provides the geometrical freedom to distort octahedra. Similarly, there are reasons to suspect that simpler physics such as positional symmetry breaking as well as magnetic symmetry breaking -- both sanctioned by single-determinant mean-field band structure view -- could also affect the effective masses in *3d oxides*. For example, oxide perovskites are known to manifest octahedral rotations and tilting [43], displacements [44], bond disproportionation [45], and Jahn-Teller distortions [46], and such local modes can couple to the electronic structure, leading to shifted band energies [30,47–52], thus possibly leading to mass renormalization. Also, whereas the PM phases in 3*d* oxides were once treated as NM [12–18] [53–55] (thus, interpreting the zero global magnetic moment as being zero on an atom-by-atom basis), more recent theories allowed for the existence of a *distribution* of different spin environments adding up to zero, constituting a *polymorphous network* that couples to electronic properties [36,56–58]. Therefore, the existence of a distribution of positional as well as local magnetic environments needs to be investigated for its ability to affect the band structure and hence the effective masses.

Thus, instead of leapfrogging from N-DFT to dynamically correlated methods such as DMFT, it would seem informative to retain the $m^*_{\text{model}}$ mass, generally used in the literature as a reference state for establishing mass enhancement, but replace $m^*_{\text{Theory}}$ in $\beta(\text{Theory/model})$ = $m^*_{\text{Theory}}/m^*_{\text{model}}$ by mean-field theory that allows for possible magnetic, orbital, and structural degrees of freedom, which could break symmetries while lowering the total energy. Such $\beta(\text{DFT/N-DFT})$ = $m^*_{\text{DFT}}/m^*_{\text{N-DFT}}$, for example, using DFT that is free from oversimplified approximations which are not an essential part of DFT, would establish which physical mechanisms contribute to mass enhancement. This might include symmetry-breaking effects routinely included in contemporary DFT calculations, such as (i) *positional symmetry breaking* such as bond disproportionation, Jahn-Teller distortions, octahedral rotations, all observed experimentally, and (ii) *magnetic symmetry breaking*, e.g., allowing spin configurations such as antiferromagnetic (AFM) and paramagnetic (PM) rather than the NM approximation. Effective allowance for (i) and (ii) also necessitates the use of XC functionals that produce correctly compact orbitals (due to closer adherence to the generalized Koopmans condition [59]) but not an overestimated orbital localization (such as in the Hartree-Fock functional), thus able to have the spatial resolution needed to 'see' atomic-scale symmetry breaking.

That symmetry breaking in approximate mean-field theory could capture events that in restricted symmetric structures would require a complex correlated treatment, has been amply illustrated in molecular systems. For example, as pointed out in 1972 by Bagus and Schaefer [60], describing a core hole state in diatomic $O_2$ while retaining the high $D_{\infty h}$ symmetry is made computationally difficult by the extreme complexity of packing the electron-electron pair correlation into a small, symmetric space. Yet breaking molecular orbital symmetry by placing the hole initially on a single atom in Hartree-Fock calculation agrees well with the experiment. Symmetry can be restored afterward [61] and often gives for localized states but small additional energy lowering. Thus, correlated methods in symmetry-restricted structures might not be the only way to describe mass enhancement. A review of the traditional role of symmetry-broken mean field vis-à-vis symmetry-unbroken correlation and that these are non-additive effects has been recently presented by Perdew *et al*. [62]

Such symmetry-broken mean-field approach can furthermore provide clear intuition as to the mechanism whereby the electronic structure (here, mass enhancement) is established. The classic picture of the face-centered cubic "empty lattice" band structure, having but the symmetry of primitive cell without interactions, as illustrated by F. Herman in 1958 [63], shows broad bands and high degeneracies with low masses. Any successive introduction of interaction terms into this empty lattice Hamiltonian (starting with the point-ion pseudopotential) would progressively remove band degeneracies, split broad bands into sets of sub-bands, and lead to mass enhancement. Examples of known modalities of symmetry breaking that are now shown to lead specifically to mass enhancement are summarized in Fig. 1 and discussed in the following sections. We find that indeed these energy-lowering, mean-field, symmetry-broken modalities (lines 2-6) can enlarge the band gap and/or contribute to more localized wavefunctions, thus leading to mass enhancements, not only for electrons but also for holes, which were previously attributed exclusively to explicitly correlated methodologies.

| Modality | | Symmetry breaking degree of freedom | Schematic | Example in this work |
|---|---|---|---|---|
| (a) Strong electronic correlation | | Electronic symmetry breaking | | |
| (b) Magnetic symmetry breaking | | Different local spin configurations | | $SrVO_3$ (PM cubic) |
| Positional symmetry breaking | (c) Octahedral rotation | Rotation angles | | $CsPbI_3$, $SrTiO_3$ (NM cubic) |
| | (d) Atom displacement | Local polarizations | | $BaTiO_3$ (AFE cubic) |
| | (e) Bond disproportionation | Different octahedral volumes | | $SrBiO_3$ (NM monoclinic) |
| | (f) Jahn-Teller distortion | Inequivalent bond lengths | | $LaMnO_3$ (AFM orthorhombic) |

**Figure 1 (1.5 columns) |** Modalities of symmetry breaking illustrated for the perovskite structure. From top to bottom: (a) Strong electronic correlation as schematically shown by the Hubbard model; (b) magnetic symmetry breaking such as the paramagnetism (PM), where the lattice sites are occupied by atoms having opposite spins without long-range order; (c) octahedral rotations allowing non-zero rotation angles; (d) atomic displacements such as the ferroelectric displacements in perovskites inducing a local polarization degree of freedom; (e) bond disproportionation allowing octahedra in perovskites to have different volumes; (f) Jahn-Teller distortions elongating the perovskite octahedron along one direction, leading to inequivalent bond lengths between the center and corner atoms.

The intuition behind this investigation originates from that the factors that control effective masses can be gleaned qualitatively from $\boldsymbol{k} \cdot \boldsymbol{p}$ perturbation theory [64],

$$\frac{1}{m^*_{n_j}} = \frac{1}{m_0} + \frac{2}{m_0^2 k^2}\left(\sum_{m \notin \{n\}} \frac{|\langle n\mathbf{0}|\mathbf{k}\cdot\mathbf{p}|m\mathbf{0}\rangle|^2}{E_{n\mathbf{0}} - E_{m\mathbf{0}}} + \epsilon_{n_j,\mathbf{k}}\right) \quad (1)$$

where $m^*_{n_j}$ is the effective mass of state $|n_j\mathbf{0}\rangle$ at band edge, **0** denotes the momentum where the band edge is located, subscript *j* is the index of degenerated wave functions, $m_0$ is the free-electron mass, $E_{n\mathbf{0}} - E_{m\mathbf{0}}$ are inter-band energy gaps, and $\epsilon_{n_j,\mathbf{k}}$ are the energy shifts from band degeneracy (zero if no degeneracy). The sum is over all eigenstates $|m\mathbf{0}\rangle$. This classic expression teaches that the effective mass in solids is generally enhanced by any effects that increase the inter-band energy gaps $\{E_{n\mathbf{0}} - E_{m\mathbf{0}}\}$ and/or reduce the wavefunction momentum matrix element $\langle n\mathbf{0}|\mathbf{k}\cdot\mathbf{p}|m\mathbf{0}\rangle$, *i.e.*, producing more compact wavefunctions.

Theoretically, DFT band energies $\{E_{n\mathbf{0}} - E_{m\mathbf{0}}\}$ do not need to correspond to any physical observable. The *total-energy difference gap*, *i.e.*, the difference between ionization potential $I = E(M-1) - E(M)$ and the electron affinity energy $A = E(M) - E(M+1)$ where M is the number of electrons, is a proxy to the quasiparticle gap. Since, in this definition, only the ground-state energies are involved, this total-energy difference gap can be calculated, in principle exactly from DFT, and can be directly compared to that measured from experiments [65–67]. Now, the quantity that is practically calculated often is the single-particle band gap $E_{CBM} - E_{VBM}$. Ref [67] has shown that such single-particle gap is equal to the total-energy difference band gap for the same exchange-correlation (XC) functional, if the XC is a non-multiplicative potential (operator is continuous) and the density change is delocalized when an electron or hole is added. This applies to LSDA, PBE, SCAN, and hybrid functionals. This is the procedure used in the current paper.

Table I lists mass enhancements calculated by symmetry-broken mean-field DFT in the present paper for six compounds, compared with experimental observations previously reported for SrVO$_3$, CsPbI$_3$, LaMnO$_3$, and SrTiO$_3$. Comparisons with previous DMFT calculations available for SrVO$_3$, LaMnO$_3$, and SrTiO$_3$ are also given in Table I. Many of the mass enhancements found by symmetry-broken DFT here are comparable in magnitude to the values suggested by correlated methods. We see that mass enhancement by symmetry breaking within mean-field DFT is not unique to open-shell *d*-electron compounds, and similar magnitudes of mass enhancements also exist for *p*-electron compounds. This further suggests that viewing the presence of mass enhancement in the considered systems as evidence for the exclusive need for strong dynamically correlated methodologies [15–24], is not a safe practice before examining the effect of positional and magnetic local environment effects on the band structure.

**Table I |** Summary of mass enhancement factors $\beta_e$ for electron and $\beta_h$ for hole, compared with experimental observations and DMFT calculations, for the compounds investigated in this work. We provide (i) the observed low-temperature low-symmetry ground states, (ii) the phases studied in this work for which mass enhancements were reported in the literature, (iii) the symmetry-restricted models used in the literature by DFT and DMFT, (iv) symmetry-broken modes found presently by DFT in the phases studied, and (v) the mass enhancements found in this work. Magnetic orders NM, AFM and PM denote nonmagnetic, antiferromagnetic and paramagnetic, respectively. Here several different DMFT values are given for comparison, because DMFT calculated mass enhancement depends on the $U$ value and the method to remove the double-counting potential. For cubic $CsPbI_3$, there are no experimental reports, to the best of our knowledge, for separate electron and hole masses; only reduced mass $m_r^* = -m_e^* m_h^* / (m_e^* + m_h^*)$ has been reported for $CsPbBrI_2$ [68]; our result is compared with such experimental reduced mass.

| Compound | Ground state | Phase studied in this work | Symmetry-restricted model (ref β=1) | Symmetry-broken mode found by DFT | | Mass enhancement |
|---|---|---|---|---|---|---|
| $SrVO_3$ | PM cubic | PM cubic | NM $Pm\bar{3}m$ | Magnetic symmetry breaking | $\beta_e$ | **DFT:** 1.5±0.1, (**exp.:** 1.3[b], 2.2[c], 2.9[c], **DMFT:** 1.8±0.2 [a]) |
| | | | | | $\beta_h$ | **DFT:** 1, (**exp.:** 1[d]) |
| $CsPbI_3$ | NM orthorhombic | NM cubic | NM $Pm\bar{3}m$ | Octahedral rotation | $\beta_e$ | **DFT:** 1.8 [†], (**exp.:** $m^*$=0.12$m_0$[†]) |
| | | | | | $\beta_h$ | **DFT:** 2.2 [†], (**exp.:** $m^*$=0.12$m_0$[†]) |
| $LaMnO_3$ | AFM orthorhombic | AFM orthorhombic | AFM $Pnma$ (no Jahn-Teller) | Jahn-Teller distortion | $\beta_e$ | **DFT:** 1.8±0.5 |
| | | | | | $\beta_h$ | **DFT:** 1.6±0.4, (**exp.:** 2.6-2.8 [e], **DMFT:** 1.3-1.7 [f]) |
| $SrBiO_3$ | NM monoclinic | NM monoclinic | NM $P2_1/n$ (no disproportionation) | Bond disproportionation | $\beta_e$ | **DFT:** 1.3±0.2 |
| | | | | | $\beta_h$ | **DFT:** 1.5±0.2 |
| $SrTiO_3$ | NM tetragonal | NM cubic | NM $Pm\bar{3}m$ | Octahedral rotation | $\beta_e$ | **DFT:** 1.1, (**exp.:** 2-3 [g], **DMFT:** 1 [g]) |
| | | | | | $\beta_h$ | **DFT:** 1.1 |
| $BaTiO_3$ | NM rhombohedral | NM cubic | NM $Pm\bar{3}m$ | Ferroelectric displacement | $\beta_e$ | **DFT:** 1.1 |
| | | | | | $\beta_h$ | **DFT:** 1 |

† DFT calculated reduced mass $m^*$=0.14$m_0$, experimentally measured reduced mass (for cubic CsPbBrI$_2$) $m^*$=0.12$m_0$ [68].
[a] Reference [11,13,14]; [b] reference [16]; [c] reference [17]; [d] estimated from reference [69]; [e] reference [70]; [f] by comparing this work with reference [53]; [g] reference [71].

## II. Approach

***Supercell model.*** To allow for inclusion of the pertinent symmetry lowering effects, the tradition of using the most economical minimal unit cell that might geometrically disallow symmetry lowering must be avoided; instead, one should use a $N_1 \times N_2 \times N_3$ replica of such minimal cell (i.e., a supercell). Table I shows for each compound (i) the observed low-symmetric ground state at low temperature, (ii) the mass-enhanced phase studied in this work, (iii) the symmetry-restricted minimal-cell model previously used in the literature. We see that the structure for which mass enhancement was reported in the experiment and studied in previous DMFT calculation is not always the ground-state structure, *e.g.*, for halide perovskites such as CsPbI$_3$ and oxide perovskite SrTiO$_3$, the ground states are orthorhombic and tetragonal, respectively, while it is the cubic phase where mass enhancements were reported [4,71]. In such cases where the phase studied is not the ground-state phase, the supercell model is constrained throughout the calculation to have the global lattice symmetry of the phase studied (*e.g.*, cubic), but the cell internal local modes are allowed to attain the values that minimize the (constrained) total energy. To assure that the relaxed atomic positions are reliable (e.g., not saddle points on the potential surface), all atoms have been initially 'nudged' by applying random atomic displacements (random for both direction and amplitude), prior to starting the process of following force minimization. The size of the supercell is increased until convergence on total energy is found; the cell size is not directly important *per se* except that certain cells cannot, by symmetry, allow symmetry breaking even if it will lower the energy. This dependence of total energy per atom on the cell size is unique to certain (polymorphous network) compounds but not in conventional materials, such as silicon or ZnSe, that do not show different total energies between supercell and minimal-cell model. Thus, larger than minimal (super) cells do not necessarily lead to disorder or mass enhancement unless energy lowering takes place.

In all magnetic calculations, the *directions* of spin moment on every site are collinear and fixed (i.e., spin flip is not allowed), while the *amplitudes* of magnetic moment are free to evolve during the total energy minimization. In principle, spins can relax during the electron self-consistency calculations to produce non-random configurations. We have confirmed that our DFT self-consistent calculations give negligible net magnetization for all AFM and PM structures (<0.001 Bohr magneton per atom). A more advanced calculation, such as the spin dynamics combined with *ab initio* molecular dynamics approach [72], could involve the freedoms of both spin direction flip and magnetic moment varying over time. It will be interesting to compare the results from this work with the ones from spin dynamics calculations.

***Band unfolding.*** Whereas the supercell approach has the advantage of allowing the incorporation of local structural and spin motifs, it has the disadvantage of producing a non-intuitive and difficult-to-analyze dense band structure in the small reciprocal-space Brillouin zone (BZ) associated with the large real-space cell dimensions. This difficulty is overcome by applying rigorous band unfolding [73–75] to the supercell band structure, producing "effective band structures" (EBS) that replace the sharp bands of ordinary band theory by spectral functions (including both coherent and incoherent components).

***Calculation of effective mass.*** We apply four models to calculate the mass and mass enhancement. (1) Deducing the mass from the mass tensor of the reciprocal for the second derivative of *E* vs ***k*** at the band edges; if the mass tensor is anisotropic ($m_1^* \neq m_2^* \neq m_3^*$), the result mass will be calculated as $m^* = (1/m_1^* + 1/m_2^* + 1/m_3^*)^{-1}$. (2) Deducing the mass from the DOS at Fermi level $m^* \propto (D(E_F))^{2/3}$. (3) Deducing the mass from the slope of bands at Fermi level, i.e., the Fermi velocity ($v_F$), as $m^* \propto 1/v_F$. (4) Deducing the mass from the bandwidth $m^* \propto 1/W$. Note that method (1) can give the absolute mass, while methods (2)-(4) are used only for the relative mass enhancement factor $\beta = m_{\text{Theory}}^*/m_{\text{model}}^*$, but not the absolute mass. We focus here on the mass enhancement factors (relative masses) rather than on the absolute value of masses.

## III. Magnetic polymorphous network in paramagnets leads to mass enhancement: SrVO$_3$

***Representation of the PM phase as a distribution of local spin environment.*** All calculations reported here are spin-polarized, allowing up and down spins. In addition, we allow for spatial spin symmetry breaking: the PM phases of 3*d* oxides are often described in the DFT literature as being *nonmagnetic,* interpreting the PM condition of globally-zero moment on an atom-by-atom basis, deducing that each atom must be nonmagnetic [12–18] [53–55]. This strong restriction does not follow from the definition of PM or from the DFT, and as was recently recognized, it leads to rather high total energy [27]. A correct description of PM entails allowing a larger (super) cell that can accommodate different local spin environments, should they lower the total (DFT) energy. For example, in a PM crystal where each magnetic ion is locally coordinated by $N$ other magnetic ions, one can have in the collinear description a distribution of local spin environments, e.g., an up-spin ion can be coordinated by $(N - m)$ up-spin ions plus $m$ down-spin ions, where $0 \leq m \leq N$; if the up and down orientations are chosen randomly (which corresponds to the high-temperature limit of PM phase), it follows the Binomial statistics, i.e., finding an ion with $m$ down-spin neighbors follows the probability function $P(m; N, 0.5) = \binom{N}{m} 0.5^N$. This model of PM local order represents a generalization of the AFM spin configuration that includes but a single local environment (e.g., up-spin site coordinated only by down-spin sites), whereas, in the PM phase, the above noted additional local environments could exist. This is accomplished in practice by borrowing an idea known from the theory of substitutional $A_x B_{1-x}$ alloys [76]: Construct an *M*-atom supercell for composition $x$ with sites occupied by A-type and B-type atoms (here, up-spin and down-spin atoms), so that the atom-atom correlation function will mimic for a finite supercell a given statistic for the infinite cell (here, random Binomial statistics) the best possible way for a *M*-atom supercell. Such "Special Quasirandom Structures" (SQS's) identify the most economical supercells for given size *M*. We currently use the *random* spin-spin correlation (corresponding to the high-temperature limit), although the use of non-random short-range order in SQS is possible [77,78].

We consider next the PM phase of SrVO$_3$ modeled by a 128-f.u. supercell (i.e., 640 atoms per cell) with collinear up and down spin configuration. Figure 2b shows in red the distribution of DFT calculated up-spin and down-spin magnetic moments, whereas the single vertical blue line shows the all-site-having-zero-spin condition in the minimal-cell NM case. For all relaxed SrVO$_3$ supercells, we have found negligible atomic displacement, consistent with the fact that the size mismatch factor revealed by the Goldschmidt factor is negligible. Comparing different spin configurations including FM, AFM, and PM phases in SrVO$_3$ (Table II in Appendix B) shows nearly degenerate total energy, suggesting that the spin ground state could be a mixture of many possible magnetic orders, i.e., in agreement with experiment [79,80] that SrVO$_3$ is a PM metal down to low temperatures and does not form magnetic order. Note that all magnetic phases show significant energy lowering when compared to the conventional NM model.

***Electron mass enhancement in SrVO$_3$.*** Figure 2a shows the N-DFT band structure (cubic unit cell containing a single f.u., no relaxation, with NM spin configuration having zero moments at all sites) giving a metallic phase, with a conduction bandwidth of 2.5 eV. This model was used for $m^*_{\mathrm{model}}$ in many previous studies [12–17] to calculate mass enhancement using, e.g., DMFT and GW theories. Figure 2c shows the spectral functions calculated in DFT from such 640-atom supercell, unfolded into the primitive BZ of SrVO$_3$. The unfolding procedure used in Figure 2c allows one to reduce the band structure complexity of a supercell, converting the sharp bands from the monomorphous case to an effective band structure having a finite band spread that depends on the distribution function used to describe the spin in PM supercell (in the current case we neglect the spin-spin short-range order, so the fuzziness may be overestimated). Figure 3 shows the evolution of the unfolded band structure as the real space supercell size increases, finding convergence.

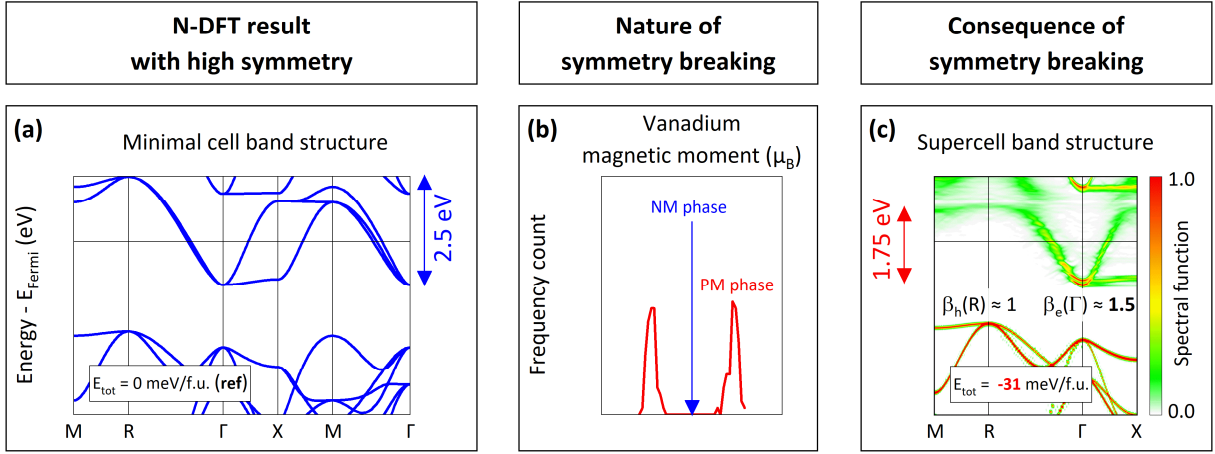

**Figure 2 (2 columns)** | Mass enhancements in the PM cubic phase of SrVO$_3$. (a) The band structure obtained from the same N-DFT restriction as in previous literature [12–17], namely a single-cell, cubic, NM SrVO$_3$ model using PBE+U (U=1.25 eV on V-d orbitals). (b) shows the distribution of spin moments in the present PM phase: Blue lines show that in the minimal-cell NM phase, all vanadium sites have zero magnetic moment, while the red curve shows that in the PM phase different vanadium sites have a distribution of different, non-zero magnetization. (c) shows the unfolded band structure when removing the minimal-cell restrictions by using instead a cubic, 128-fu PM supercell SrVO$_3$ with the same PBE+U method. Masses in (c) are calculated via DOS at Fermi level (which gives $\beta_e$=1.4-1.6 and $\beta_h$=1; the subscript *e* and *h* denote the electron and hole mass enhancements; uncertainty is due to the variation of DOS nearby the Fermi level), second derivative of *E* vs *k* (which gives $\beta_e$=1.46 and $\beta_h$=1), and bandwidth (which gives $\beta_e$=1.43 and $\beta_h$=1). The vertical arrows in (a) and (c) show the bandwidths. The same PBE+U method has been applied for all SrVO$_3$ calculations.

The significant result is that the conduction band in the PM phase, allowed to have a distribution of local spin motifs, is narrowed relative to the minimal-cell NM case, from the 2.5 eV (Figure 2a) to 1.75 eV (Figure 2c). This leads to electron mass enhancement factor *($\beta_e$)* in the PM supercell $\beta_e$(DFT/N-DFT)=1.43. Note that different definitions of effective mass give somewhat different results: The bandwidth mass enhancement of 1.43 can be compared with the density-of-states mass enhancement 1.4-1.6 at Fermi level, while the second derivative of *E* vs *k* at the conduction band edge gives 1.46. Note that we have not attempted to fit the result by adjusting U, although the choice of more localizing XC functional can increase the enhancement factor. We note that whereas the values obtained depend somewhat on the definition of effective mass used, the lattice constant (here we used

a=3.83 Å), and the *U* value, allowing for polymorphous spin configuration, leads in all cases to an enhancement factor of 1.5±0.1. These calculated mass enhancements are comparable to the experimentally measured factor $\beta_e$(exptl/N-DFT)≈1.8±0.2 [11,13,14], while smaller than the enhancement factors from DMFT, *e.g.*, $\beta_e$(DMFT/N-DFT)=2.9 [17] (using a much larger U=5.5 eV that narrows bands further) and from GW+DMFT $\beta_e$=1.3 [16], 2.2 [17] (where the two values correspond to different versions of accounting approximately for the double-counting error in GW+DMFT).

*Hole mass enhancement in SrVO₃.* The DFT calculations naturally provide all bands with equal approximations, in particular, both the electron conduction band (mainly V-*d* orbitals) and the valence hole band (mainly O-*p* orbitals). We find no mass enhancement for hole states in the principal valence band, consistent with the fact that the spin configuration in the O *p*-like principal valence band corresponds to a closed electronic shell and negligible magnetic moment that show no distribution of motifs. We will see later that in ABX₃ perovskites, where the local environment is made of a distribution of *positional* motifs rather than from spin motifs, there will be both electron and hole mass enhancements.

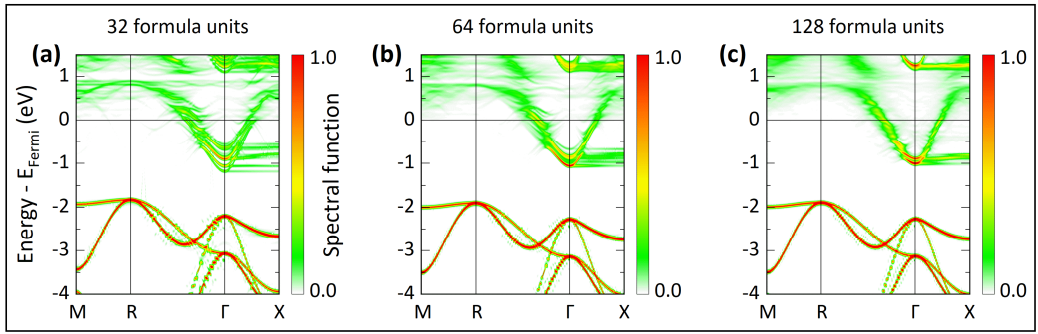

**Figure 3 (2 columns)** | Convergence of mass enhancement and band structure of SrVO₃ PM vs. supercell size increasing from 32 formula units (160 atoms), to 64 f.u. (320 atoms), to 128 f.u. (640 atoms). All supercells are generated by using the spin SQS method. Note band narrowing convergence.

*Analysis of the contributing factors to spin-induced mass enhancement in SrVO₃.* The real-space symmetry-broken supercell approach provides for an intuitive understanding of the results. The degree of freedom within our PM supercell is the local spin configuration (as we have found that the positional relaxation is negligible in this system). A local spin motif consists at the first order of a central 3*d* atom and its first-shell 3*d* (next nearest) neighbors. Whereas in the case of the minimal-cell NM model, each and every motif has zero spin, and in the case of AFM order, each motif has maximum dissimilarity between the spin of the central atom and the spins of its coordination shell (*e.g.*, the AFM-G like local motif when the central atom is up-spin, while all its neighbors are all down-spin), in our model of the PM phase, each spin can have a local distribution of spins, including the case of maximum similarity (*e.g.*, the FM-like local spin motif when the central atom is up-spin, and so are all its neighbors), maximum dissimilarity, or any configurations in between. Figure 4(a)-(c) shows the *random* (high-temperature limit) statistical weight for each spin motif. Each local spin motif might have its unique, projected local density of states. The vertical arrows in Figure 4(d-i) indicate that each local spin motif contributes differently to the conduction bandwidth, whereas the reference minimal-cell result using the NM model (shown in Figure 4(d)) has considerably wider DOS. Figure 4(e-i) shows that the locally AFM-G-like vanadium sites with max spin dissimilarity with their neighbors have the most *compressed* DOS (Figure 4(i); smallest range in energy and the highest peak in DOS, indicating enhanced electron mass), while the locally FM-like sites with max spin similarity with the neighbors have the most *expanded* DOS (Figure 4(e); similar to the bandwidth in the NM model in Figure 4(d), hence not contributing to mass enhancement). Figure 4 suggests that the mass enhancement depends on spin short-range order (SRO), *i.e.*, the spin configuration at the center site and the neighbors, through the statistical weight for each spin motif (Figure 4b): an AFM-like, anti-clustering SRO can lead to a larger enhancement coefficient $\beta$, while a FM-like, clustering SRO can reduce $\beta$. This analysis teaches that the existence of a polymorphous distribution of spin-polarization motifs with their attendant, different local density of states contributing differently to the total DOS creates the possibility of *spin-induced mass enhancement*.

As discussed above, static DFT calculations permitting symmetry breaking and ensuing creation of local spin motifs naturally show mass enhancement. The scope of the current calculation of E-vs-k dispersion does not extend, however, to fully model ARPES spectra, including ARPES matrix element effects [81,82], or lifetime effects [83,84] associated with the observed sharpening of the ARPES states as they approach the Fermi level [12–14]. Description of lifetime-broadening generally requires a time-dependent dynamics analysis, *e.g.*, spin dynamics combined with molecular dynamics, possible in the mean-field DFT framework [72].

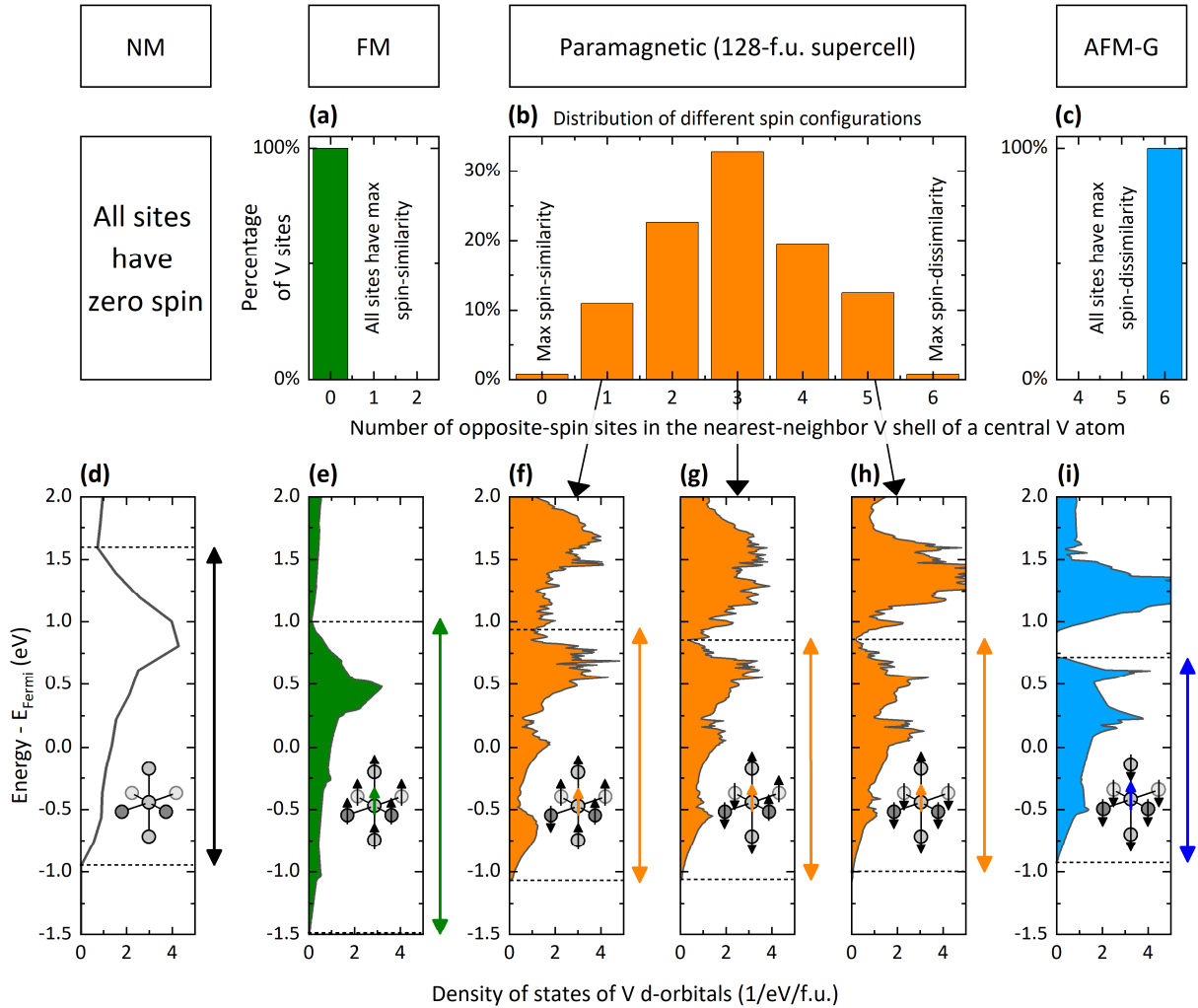

**Figure 4 (2 columns) |** Total DOS of cubic SrVO$_3$ (the same PBE+U method) as a weighted superposition of the partial DOS (PDOS) of the local spin motifs. Upper panel -- the weight of local spin configuration in: (a) Statistical weights of FM primitive cell (all the first-neighbor vanadium have the *same* spin direction as the center vanadium, see insert in (a)); (b) distribution of statistical weights of different spin configurations in the cubic PM supercell; (c) statistical weights of AFM-G double perovskite cell (all the first neighbor vanadium have the *opposite* spin direction to the center vanadium, see insert in (c)). Lower panel -- the vanadium *d*-PDOS in different spin configurations: (d) NM primitive cell in which each all site has zero spin; (e) FM motif; (f)-(h) different local spin motifs in PM phase, and (i) AFM-G motif. The dash lines and vertical arrows in (d)-(i) show the bandwidths of *d*-orbital for guide of eyes. Note that in (d)-(i) we only show the *d*-PDOS of the central vanadium atoms (neighbors are not included to avoid multiple counting).

# IV. Positional symmetry breaking causes coupling of the electronic bands leading to mass enhancement.

## A. Octahedral rotation enhances masses in *s-p* halide perovskite CsPbI$_3$

Octahedral tilting (rotations) [43] have been known for long to exist in perovskites; here, we point out that such observed local modes can cause mass enhancement. The perovskite structure consists of corner-sharing octahedra, that allow octahedral rotation and tilting. The classic atomic size mismatch between the A and B sublattices in ABX$_3$ drives static octahedral tilting and rotations, as was recognized already in 1926 [85] and verified by modern PDF measurements [86] as well as by DFT total energy minimization [52,87–90]. This kind of deformation derives from classical atomic size mismatch and therefore exists even in close-shell *s-p* electron compounds such as halide perovskites [52,88,89]. Octahedral tilting (rotation) effects have been investigated in lead and tin halide perovskites for the low temperature *tetragonal and orthorhombic phases*. However, for the *cubic phase,* it has been generally assumed [30–33] that because of its XRD designation as the $Pm\bar{3}m$ structure with a single formula unit per cell, such tilting is disallowed by symmetry hence absent. Here we point out that (i) a minimal 2×2×2 cubic supercell is needed to reveal tilting, (ii) *single* tilting modes [43] already change the mass, and (iii) larger supercells such as 4×4×4 can reveal *multi-mode* tilting, showing further lowering of the total energy relative to the monomorphous $Pm\bar{3}m$ cubic model. Such multi modes significantly affect mass enhancement. Although technically this can be posed as an electron-phonon effect, given that the pattern of tilting and atomic displacements are complex and imply the participation of a significant number of phonons, we do not use the phonon terminology; instead, we explicitly (and non-perturbatively) minimize the generally anharmonic DFT energy surface with respect to all atomic positions, dislodge atoms from metastable local positions to stable positions, then calculate the ensuing band structure and masses.

*Qualitative analysis of how octahedral tilting affects band edge energies in CsPbI$_3$.* The coupling between octahedral tilting and electronic structure in low-temperature phases has been discussed in oxide [5] and halide perovskites [30] as a band-gap-tuning mechanism. Here as schematically shown in Figure 5, we demonstrate how the octahedral rotation in *s-p* bonded compound CsPbI$_3$ affects band edge states. In cubic CsPbI$_3$ without distortion, the valence band maximum (VBM) is an antibonding state formed from the Pb-*s* and I-*p* orbitals, while the conduction band minimum (CBM) is weakly antibonding state of Pb-*p* and I-*p* orbitals. Allowing rotations of the (PbI$_6$) octahedron will weaken both the *p-p* and *s-p* bonding between Pb and I. Consequently, being antibonding states, *both CBM and VBM* will move to *lower* energies. However, because the VBM is more sensitive to octahedral deformation (being composed of inner-shell 6*s*-3*p* antibonding) than the CBM (being composed of outer shell 6*p*-3*p* antibonding) [91], the VBM will move further than CBM, leading to a larger band gap due to rotation (Figure 5, right panel).

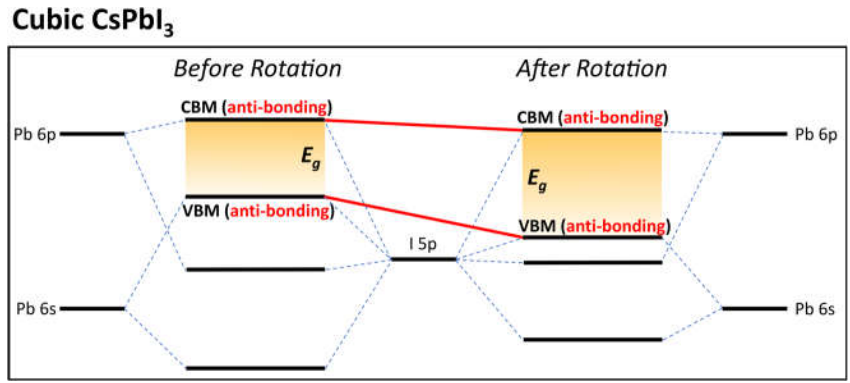

**Figure 5 (1.5 columns)** | Energy level diagram for CsPbI$_3$, before and after octahedral rotation. The red solid lines show the trends of band gap change.

***Model DFT calculations on the effect of frozen rotations on mass enhancement in CsPbI$_3$.*** The results of the simple model of Figure 5 are then validated by DFT calculations of small (8-f.u.) model supercells where we artificially impose given octahedral rotation angles, followed by band unfolding to the single-cell cubic primitive Brillouin zone. The unfolded band structure is shown in Figure 6. The imposed octahedral rotation here is a$^+$a$^-$a$^-$ mode (Glazer notation [43]) or $M_2^+ \oplus R_5^-$ (irreducible representation from S. C. Miller and W. F. Love [92]). It can be seen that the imposed octahedral rotations can affect the curvature of both valence and conduction bands at the band edges. Choosing the band gap and effective masses of the zero-rotated structures as the reference, as shown in Figure 6, one can see that under an uniformed arbitrary rotation of (10°, 10°, 10°), the band gap of CsPbI$_3$ increases by 0.52 eV (~40%), and the electron and hole masses are enhanced by 77% and 113%, respectively.

CsPbI$_3$ (nonmagnetic cubic phase + artificial tilting, PBE)

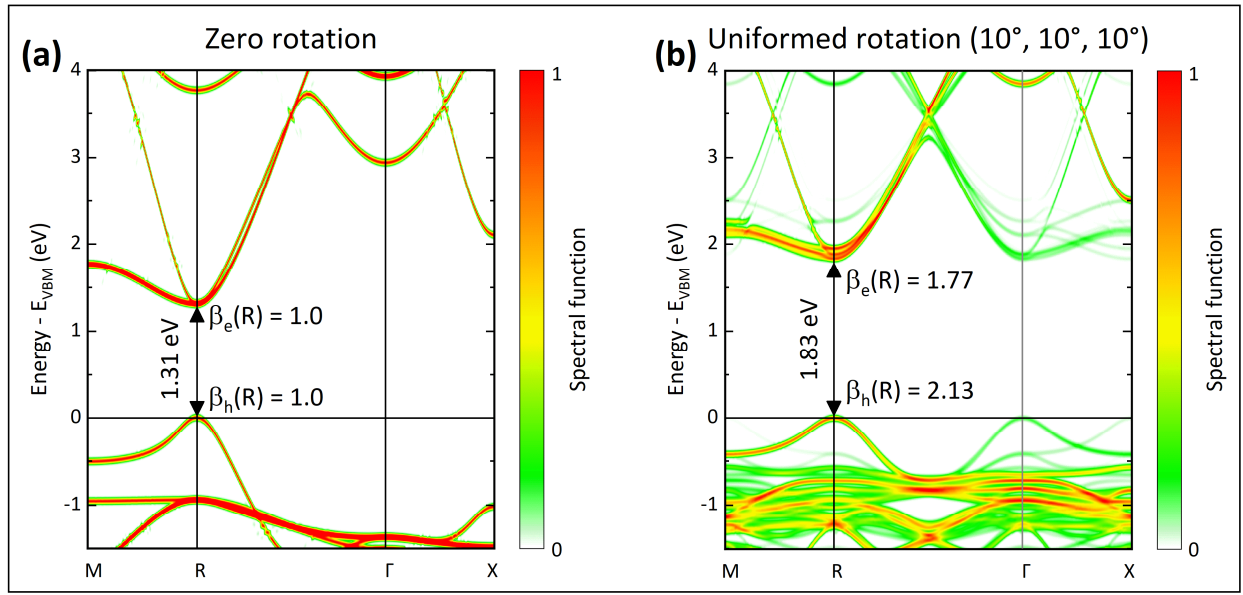

**Figure 6 (2 columns)** | The unfolded spectral function (EBS) in the cubic primitive Brillouin zone, before (a) and after (b) a uniformed (10°, 10°, 10°) a$^+$a$^-$a$^-$ rotation for cubic CsPbI$_3$. Both (a) and (b) are calculated using the 8-f.u. supercell. Note that EBS shown in (a) is identical to the band structure obtained from a minimal-cell cubic model because (a) has no atomic distortion. Band gap is 1.31 eV in (a) and 1.83 eV in (b). Taking effective masses of CBM and VBM in (a) as references, the mass enhancement factors in (b) are β$_e$=1.77 (for electron, counting all 3 states near CBM) and β$_h$=2.13 (for hole), respectively. All masses come from band dispersion (second derivative of E vs k).

*Full supercell calculation of rotation-induced mass enhancement in CsPbI$_3$.* Having clarified the effect of classical rotation on the band structure by the model (Figure 5) and validated it by DFT (Figure 6), we next study a large supercell with optimized rotating geometries in the cubic phase of CsPbI$_3$. Recall that the rotations discussed here are not thermal, but, in fact, they are energy-lowering distortions derived by the nature of the chemical bonds (here, steric effects) even at low temperatures. Thus, we obtain these deformations by minimization of the DFT internal energy. But this requires that we allow a larger than the minimal unit cell so that rotations can be accommodated geometrically. The electronic structure of a single-cell cubic ($Pm\bar{3}m$) model (monomorphous models) is shown in Figure 7(a). Restricted by the small size and periodic boundary condition, such structure cannot accommodate rotations (as shown by the blue lines in Figure 7(b)). We avoid such restrictive assumption using a supercell representation (32-f.u. supercell) instead. We perform a *constrained minimization* of the internal energy (T=0) of the cubic phase that retains the macroscopic cubic supercell shape (or else the minimization will converge to the ground-state orthorhombic or tetragonal structures that are not the subject of the current study). At the same time, we allow all cell-internal degrees of freedom to locally adjust to find the structure with the lowest total energy. This is done by using a set of random initial nudging, so as to dislocate atoms from possible local minima.

It has been shown in our recent study [36] that for lead halide perovskites with organic molecules on the A-site, such supercell representation following the constrained DFT total-energy minimization explains various anomalies in the cubic phase, where the minimal-cell model disagrees with experimental observation. This includes a close agreement with the measured pair distribution function, significant increase in the band gap, and dielectric constant. For the cubic $CsPbI_3$ supercell, we have found that when abandoning the restriction of the minimal-cell model, the atomic configuration of cubic phase that gives the lowest total energy of the cubic phase (the supercell out-shape has been constrained to be cubic), is not the single-f.u. $Pm\bar{3}m$ cubic model, but a supercell representation for the cubic phase with many local octahedral tilting. The many local tilting in the cubic supercell cannot be written using the simple Glazer notations but must involving the complex notations [43], and cannot be reduced into smaller cell models. We find that: (i) the rotation angles are distributed among 5-13° (red lines in Figure 7(b)(e)); (ii) the supercells have lower total energy (-124.4 meV/f.u.) compared with the single-cell model. Recall that during the DFT calculation, we used the equivalent k-point mesh for all cells (12×12×12 Γ-center k-point mesh in primitive cubic BZ) and a total-energy tolerance of $10^{-8}$ eV/atom, we, therefore, suggest that the -124.4 meV/f.u. energy lowering is robust. Furthermore, (iii) we find, as expected from the simple model of Figure 5, a significant band gap increase and thus mass enhancement factors: the band gap increases from 1.31 eV in the monomorphous single-cell model to 1.85 eV in the polymorphous supercell model (the measured band gap is $E_g$=1.73 eV [30]) leading to mass enhancements $β_e$=1.8 and $β_h$=2.2 for electrons and holes, respectively (Figure 7(a)(c)). Although, as far as we know, the effective mass has not been reported from experiments for cubic $CsPbI_3$, the reduced mass ($m^* = -m_e^* m_h^* / (m_e^* + m_h^*)$) for cubic $CsPbBrI_2$ has been measured [68] as $m^*$=0.12±0.01 $m_0$, which is close to our prediction (masses are calculated via the second derivative of E vs k) here $m^*$=0.14$m_0$ using symmetry-broken DFT. Recall that neglecting the distortions gives a mass 1.8 times smaller. We conclude that semiclassical octahedral rotations in $ABX_3$ can derive quantum mechanical band gap increases and significant mass enhancements. The same physics is expected to contribute to oxides; the magnitude of the effect would naturally depend on the extent of rotations and the response of the host crystal to deformations.

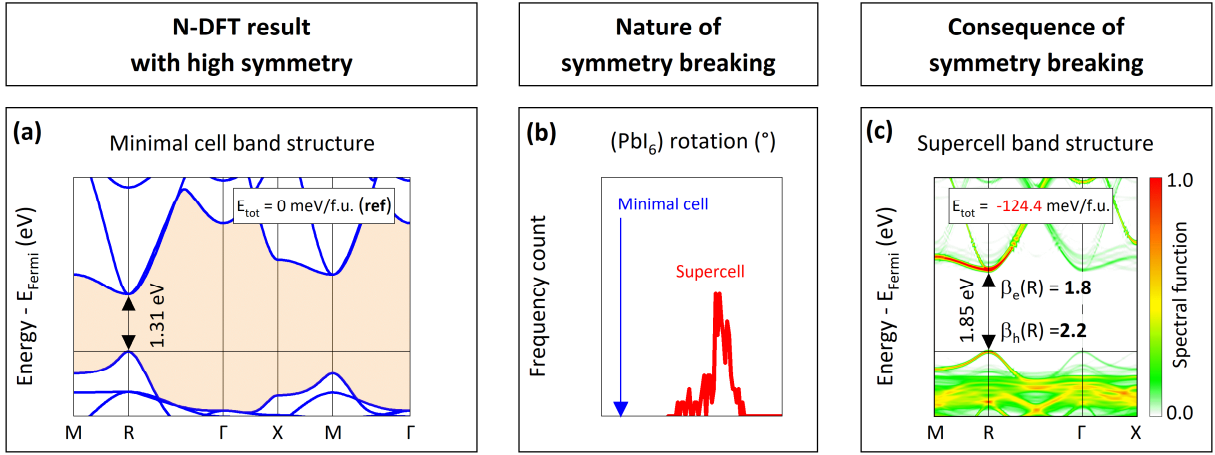

**Figure 7 (2 columns)** | Mass enhancements in cubic CsPbI$_3$. (a) The band structure from the same N-DFT restriction as in previous literature [30–33], namely a single-cell cubic model, using PBE functional. (b) gives the distribution of octahedral rotation angles. The blue arrow in (b) shows that in the minimal cell model all octahedra have zero rotation, while the red curve shows that in supercell different octahedra have a distribution of different, non-zero rotations. (c) shows the unfolded band structure when removing the minimal-cell restrictions by using instead a cubic, 32-f.u. supercell CsPbI$_3$ with the same PBE method. Masses in (c) are calculated via the second derivative of E vs k, which gives $\beta_e$=1.8 and $\beta_h$=2.2. To the best of our knowledge, the experimental measurement for separate electron and hole masses for cubic CsPbI$_3$ has not been reported yet; while for cubic CsPbBrI$_2$ the reduced mass $m^* = -m_e^* m_h^* / (m_e^* + m_h^*)$ has been reported as 0.12$m_0$, very similar to the reduced mass calculated from (c) which is 0.14$m_0$, but 1.8 times heavier than the reduced mass from (a) which is 0.07$m_0$.

## B. Jahn-Teller-like $Q_2^+$ distortion enhances masses in LaMnO$_3$

To draw the analogy between mass enhancement in *s-p* bonding perovskites (previous section) and the better-known effect in *d*-electron perovskites, we treat next the compound LaMnO$_3$. The observed positional symmetry breaking in orthorhombic LaMnO$_3$ is a pseudo-Jahn-Teller distortion, leading to inequivalent Mn-O bond lengths in MnO$_6$ octahedron. It has been previously argued that such Jahn-Teller distortion is specifically attributed to dynamic correlations [53], and that mean-field DFT fails to predict the ensuing structural or electronic properties [93]. Previous DFT+U calculations reproducing the Jahn-Teller distortion in the orthorhombic LaMnO$_3$ [94–96] gave perhaps the impression that the presence of +U in DFT implies the same correlation role as +U in the Hubbard Hamiltonian. More recently, Varignon et al. [97] showed that DFT without U is already enough to capture such distortions, provided that a more accurate exchange-correlation functional was used. It was also clarified [98] that proper Jahn-Teller distortion is associated with degeneracy removal (i.e., electronic instability such as the $Q_2^-$ mode [99]), whereas the $Q_2^+$ mode [99] originates from classic steric effects (the Goldschmidt tolerance) that can be classified as *pseudo*-Jahn-Teller distortions. Here, we show that such modes cause mass enhancement.

Figure 8(c)(d) show the atomic structure and band structure of orthorhombic AFM cell with $Q_2^+$ deformation taken from previous DFT calculation [98]; while Figure 8(a)(b) show the atomic structure and band structure of the same AFM cell (the same AFM order, the same octahedral rotation and octahedral volume) as (a)(b) but only removing the $Q_2^+$ deformation. Such $Q_2^+$-free model (Figure 8(a)(b)) has been used as the DFT model in previous studies, giving metallic behavior [53,100,101]. Both cases are calculated using SCAN functional. The band structure lacking the $Q_2^+$ mode (b) is metallic, while the band structure with $Q_2^+$ mode (d) is gapped. Using (b) as the reference and considering the bandwidths of Mn-$d$ e$_g$ bands in (d), the bandwidth enhancement factors is $\beta_e$=1.8±0.5 and $\beta_h$=1.6±0.4, for electrons and holes, respectively; the uncertainty is due to that bandwidths along different k paths give different enhancement factors. The enhancement factor obtained by DMFT with $Q_2^+$ deformation is $\beta_h$=1.3-1.7 (values are extracted by comparing this work with results reported in DMFT reference [53]), and comparable to experimental observation of $\beta_h$=2.6-2.8 in La$_{0.6}$Sr$_{0.4}$MnO$_3$ [70]. We conclude that pseudo-Jahn-Teller distortion $Q_2^+$ captured by mean-field DFT is capable of producing significant mass renormalizations, even though on this case both DFT and DMFT give smaller renormalization than what was measured, which may be due to the A-site alloy effect in the measured sample La$_{0.6}$Sr$_{0.4}$MnO$_3$ [70].

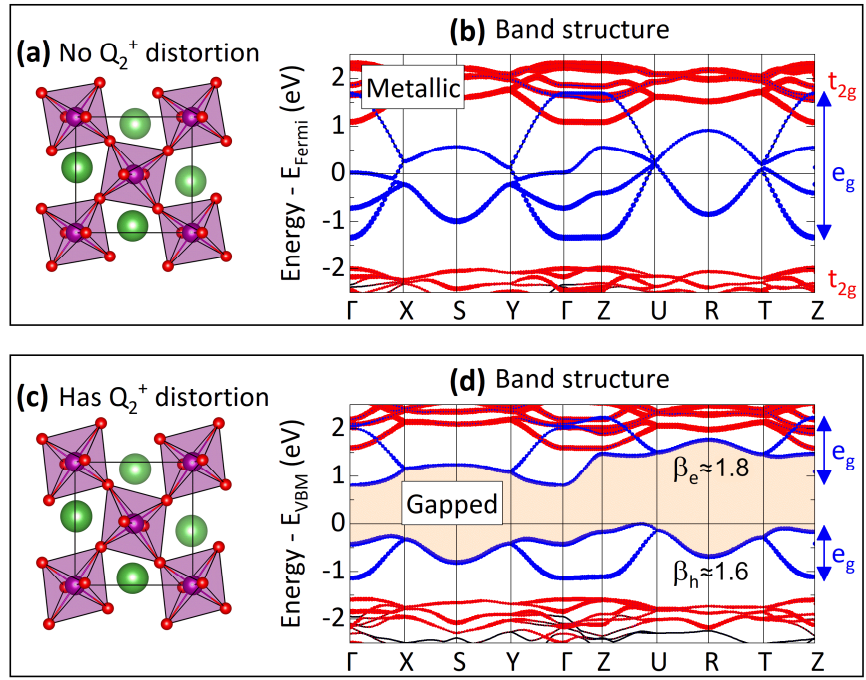

**Figure 8 (1.5 columns) |** Mass enhancements in AFM orthorhombic phase of LaMnO$_3$. (a) shows the atomic structure and (b) gives the band structure with no Jahn-Teller-like $Q_2^+$ distortion, using SCAN functional. The system is metallic and has a broad Mn $e_g$ band (bandwidth is denoted by the blue arrow on the right side of (b)) crossing the Fermi level. (c) shows the atomic structure and the (d) gives the band structure when considering the correct $Q_2^+$ distortion with the same SCAN functional, where the gap opens between the two Mn $e_g$ bands (bandwidths are denoted by blue arrows on the right side of (d)). Red and blue circle symbols in (b) and (d) denote the orbital projections of Mn-$d$ $t_{2g}$ and $e_g$ orbitals, respectively; other orbitals are not shown here; the size of the circle is proportional to the composition of orbital. Mass enhancements in (d) with respect to (b) are calculated via the bandwidths of Mn $e_g$ bands, which give $\beta_e$=1.8±0.5 and $\beta_h$=1.6±0.4 (uncertainty is due to that bandwidths along different k paths give different enhancement factors).

### C. Bond disproportionation enhances masses in SrBiO$_3$

Bond disproportionation in perovskites corresponds to the spontaneous transformation of two equal octahedra into two inequivalent octahedra, also known as octahedral breathing distortion, belonging to $M_1^+$ or $R_1^+$ mode [102]. A cell model of ABX$_3$ containing a single formula unit allows obviously but a single volume for all (BX$_6$) octahedra, while some compounds, e.g., SrBiO$_3$ and BaBiO$_3$ [29] prefer bond disproportionation on B ions, appearing as some octahedra dilate while others contract, eventually leading to multiple local environments instead of a single local environment. Total energy calculations [103] show that this disproportionation is energy lowering, not a transition between two phases.

We choose SrBiO$_3$ as an example to investigate such bond disproportionation effect on effective masses. SrBiO$_3$ is known to be insulating in its low-temperature monoclinic phase with a disproportionate $R_1^+$ distortion [104]. The monoclinic phase shows tilting $M_3^+ \oplus R_4^+$ mode (Glazer notation a$^+$b$^-$c$^-$), which could also contribute to the mass enhancement. To isolate the contribution of $R_1^+$ disproportionation from tilting, we apply here a three-level model: (1) We start from level-1 model, which is minimal-cell cubic, $Pm\overline{3}m$ structure, then we (2) apply *tilting* $M_3^+ \oplus R_4^+$ mode to construct level-2 a monoclinic ($P2_1/n$) structure without disproportionation, and finally (3) level-3 model uses the experimentally observed SrBiO$_3$ monoclinic phase (also $P2_1/n$) with both tilting (the same amplitude as in level-2) and disproportionation. The atomic structures, together with the band structures using the PBE functional + SOC effect for such three levels, are shown in Figure 9.

(1) Level-1 ($Pm\overline{3}m$ cubic; Figure 9(a,b)) is a p-type degenerate metal, as its Fermi level crosses its wide (as denoted by the blue arrow on the right side of Figure 9(b)), principal valence band made of O-*p* orbitals. The total DFT energy of level-1 is extremely high (+980 meV/f.u. above the convex hull), clarifying that it is not a low-temperature ground state. (2) Level-2 monoclinic phase without disproportionating $R_1^+$ distortion (Figure 9(c,d)) shows a more compact O-*p* valence band, however, it is still a p-type gapped metal, i.e., the octahedral tilting cannot open the gap. The total DFT energy of level-2 is 71 meV/f.u. above the convex hull. Finally, (3) level-3 the monoclinic phase with disproportionate $R_1^+$ distortion (experimental structure; Figure 9(e,f)) shows semiconducting behavior with a 0.26 eV gap between the two split O-*p* bands, a splitting induced by the bond disproportionation. Level-3 is at on the convex hull (i.e., the ground state). Nevertheless, as seen in Figure 9, the disproportionating $R_1^+$ distortion is the key to band gap opening, therefore, the most interesting mass enhancement is the one from level-2 to level-3 $\beta(L3:L2)$. Considering the bandwidths of O-*p* bands in Figure 9(d) and (f), if using Figure 9(d) as the reference, the bandwidth-related masses in Figure 9(f) have the enhancement factors of $\beta_e(L3:L2)$=1.3±0.2 and $\beta_h(L3:L2)$=1.5±0.2; the uncertainty is due to that bandwidths along different k paths give different enhancement factors. We conclude that disproportionation symmetry breaking, an effect that exists both in *s-p* perovskites as well as in *d*-electron perovskites, is capable of significant mass enhancement.

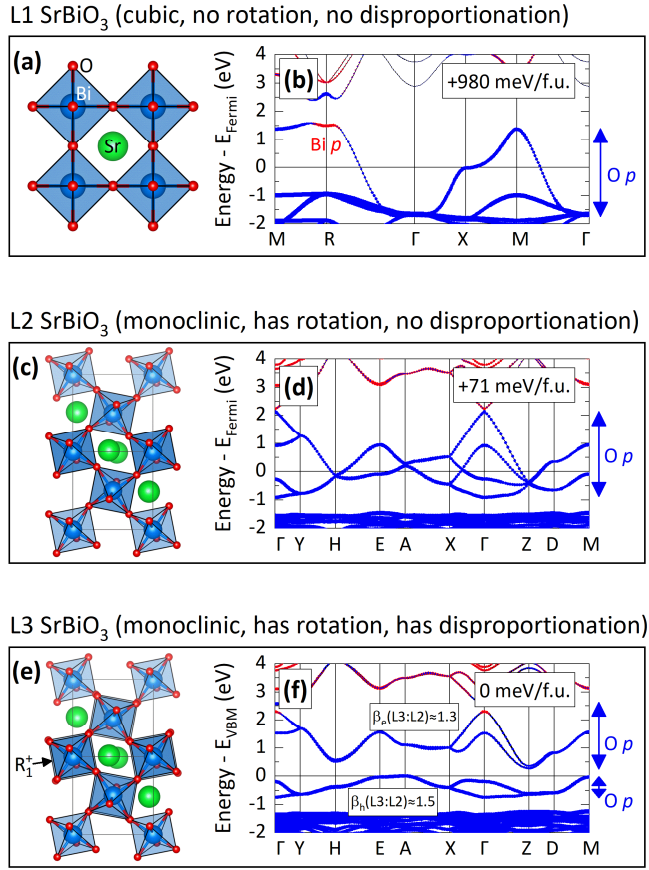

**Figure 9 (1 column) |** Mass enhancements in SrBiO$_3$. (a) shows the atomic structure and (b) gives the band structure from the $Pm\bar{3}m$ phase, using PBE functional (level-1, L1). The system is metallic and has a broad O-$p$ band (blue arrow on the right side of (b)). (c) shows the atomic structure and the (d) gives the band structure when allowing the observed (monoclinic) octahedral rotations, but no disproportionate $R_1^+$ distortion, with the same PBE functional (level-2, L2), where the O-$p$ bands become more compact, but the system stays metallic. (e) shows the atomic structure and the (f) gives the band structure when allowing all experimentally observed distortions, with the same PBE functional (level-3, L3), which leads to the O-$p$ bands splitting and hence a band gap opening. The $R_1^+$ distortion can be seen by the slightly different octahedral volumes in (e) (black arrow). Red and blue circle symbols in (b) (d) and (f) denote the orbital projections of Bi-$p$ and O-$p$ orbitals, respectively; other orbitals are not shown here; the size of the circle is proportional to the composition of orbital. Mass enhancements in (f) with respect to (d), i.e., $\beta(L3:L2)$, are calculated via the bandwidths of O-$p$ bands, which give $\beta_e$=1.3±0.2 and $\beta_h$=1.5±0.2; the uncertainty is due to that bandwidths along different k paths give different enhancement factors. The same PBE method with spin-orbit coupling has been applied for all SrBiO$_3$ calculations.

## V. Not all positional symmetry breakings lead to significant mass enhancement.

The examples shown in section IV A, B, and C indicate cases where symmetry breakings result in large energy-lowering and they couple significantly to the electronic manifold, altering its band structure, including effective mass enhancement. There are, however, cases where such deformations are small, or even if not small, they might couple only weakly to the electronic states that form the band edges, i.e., small deformation potentials, leading to small or negligible mass enhancement. We next illustrate two such examples.

## A. Weak octahedral rotations in intrinsic SrTiO₃ cause negligible mass enhancement.

Undoped SrTiO₃ is a nonmagnetic insulator. Positional distortions, such as the octahedral tilting in SrTiO₃ (also known as the antiferrodistortive displacement, AFD), have been noted in the low-symmetry tetragonal phase [105–107], but *absent* in the cubic phase [107–110]. Some observations of distortion in the cubic phase were then attributed to extrinsic factors such as strain [111–114], defects [115], impurities [116], interface [117], or thermal effect [118]. Indeed, AFD is not expected theoretically in the cubic phase unless one abandons the conventional single formula unit monomorphous description of the $Pm\bar{3}m$ XRD model and allows tilting degree of freedom in a supercell description of the cubic phase. As Table I indicates, here, we focus on the properties of the cubic phase, not the low-temperature ground state, performing a minimization of the cubically constrained total energy as a function of the cell internal atomic positions.

When doped n-type, *e.g.*, SrTiO₃:Nb or SrTiO₃:La at doping concentration of 0.01-0.05 electron/f.u., one observes (i) the formation of a low dispersion (heavy mass) occupied *in-gap* states [119–121], as well as (ii) a Fermi level inside the broad (light mass) principal conduction band. Using plasma frequency as a measure for electron mass $\omega^2 = n_c/m^*$, where $n_c$ is the free carrier concentration, large values of electron mass enhancement of $\beta_e$=2-3 [71] were reported. We find that even if removing the single-cell restrictions used in previous studies (i.e., $Pm\bar{3}m$ minimal cell with a single f.u., as shown in Figure 10(a)) by constructing a supercell (here, 64-f.u.) and relaxing all internal atomic positions, the cubic phase SrTiO₃ shows only small octahedral rotations around 3-4 degrees (shown in Figure 10(b)), a small total energy lowering of -4.4 meV/f.u., and negligible enhancements for both electron and hole masses (Figure 10(c); $\beta_e$≈1.1 and $\beta_h$≈1.1, both calculated from the second derivative of E vs k). Thus, our calculated value pertaining to the principle conduction band (not in-gap polaron like state) $\beta_e$≈1.1 indicates negligible enhancement. In the doping concentration 0.05 electron/f.u., both DMFT [108] and N-DFT [108] calculations predict similar electron mass, indicating no mass enhancement by strong electron correlations. It is not clear if the observed substantial electron mass enhancement is related to the presence of polaron states [(i) above] or is intrinsic [(ii) above, due to electron doping [115,116,122] of main conduction band].

Indeed, the plasma frequency mass enhancement $\beta = \omega_{model}^2/\omega_{exp}^2$ may be affected by doping compensation, reducing the effective concentration of free carriers. The experimental values of mass enhancement in SrTiO$_3$ appear to require clarification.

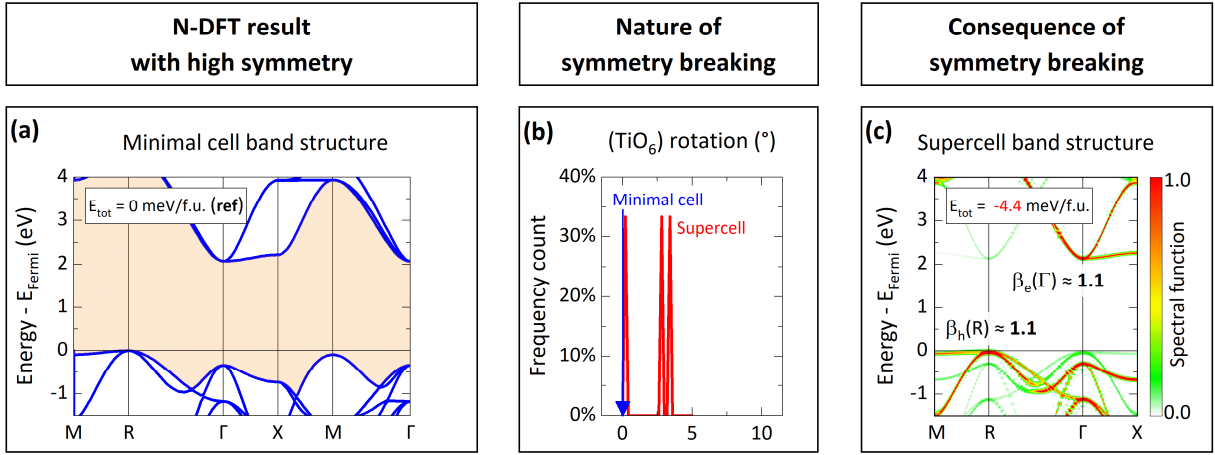

**Figure 10 (2 columns) | Mass enhancement in the nonmagnetic cubic phase of SrTiO$_3$.** (a) shows the band structure (from the same N-DFT restriction as in previous literature [108–110], namely a single-cell, cubic, NM SrTiO$_3$ model, using SCAN functional. (b) shows the distribution of octahedral rotation angles. The blue arrow in (b) shows that in the minimal cell model all octahedra have zero rotation, while the red curve shows that in supercell different octahedra have a distribution of tiny rotations around 3-4 degrees. (c) shows the unfolded band structure when removing the minimal-cell restrictions by using instead a cubic, 64-fu NM supercell SrTiO$_3$ with the same SCAN method. Masses in (c) are calculated via the second derivative of $E$ vs $\mathbf{k}$, which gives $\beta_e \approx 1.1$ and $\beta_h \approx 1.1$.

### B. Antiferroelectric and paraelectric displacements in BaTiO$_3$ have negligible effects on masses

Another well-known case where local atomic displacement occur involves ferroelectric (FE) compounds, often having paraelectric (PE) and antiferroelectric (AFE) phases. BaTiO$_3$ is one of the first-found FE perovskite compounds [123–125], where the ferroelectricity is induced by the off-center displacement of the Ti atom in (TiO$_6$) octahedron. BaTiO$_3$ experiences a complex phase transition, from rhombohedral (R, <180 K) to orthorhombic (O, <280 K), to tetragonal (T, <400 K), to cubic (C) [126]. While its R, O, and T phases all show ferroelectricity, the high-T cubic phase shows no net ferroelectricity. Therefore, it has been argued that such cubic phase has no Ti atom off-center displacement in any octahedra (i.e., a non-electric (NE) phase) and can be represented by a minimal cubic cell model [127–129]. However, recent investigations show that such minimal-cell, NE cubic model cannot

25 / 40

explain the Raman and X-ray fine structure (XAFS) observations, and the cubic phase could be instead an AFE phase [44] or a PE phase [130]. Here we aim to study if the AFE and PE nature in cubic BaTiO$_3$ can have effects on its electronic properties, such as band gap and effective mass.

The AFE phase is mimicked by an 8-f.u. supercell, constraining its lattice vectors to the macroscopically observed cubic structure while relaxing all cell-internal atoms. Figure 11(a) shows the band structure of the non-electric model (single-f.u., NE cubic model) using the SCAN functional. The difference between the atomic positions in the NE monomorphous model and the AFE polymorphous supercell is plotted in Figure 11(b): The monomorphous cell has no Ti-atom displacement (ΔR$_i$=0 for every Ti), while the AFE supercell shows a unique displacement pattern, where the eight Ti atoms move along the eight <111> directions (i.e., [111], [11-1], [1-1-1], …); although the AFE supercell has a zero net polarization (<Δ**R**>=0), the local polarization on each Ti site is non-zero and as large as 0.13 Å (<|Δ**R**|>=0.13 Å). Other distortions (rotations, Jahn-Teller distortion, etc.) have all been found to be negligible in the AFE supercell. This AFE displacement pattern agrees well with the previous theory [44]. Figure 11(c) shows the unfolded band structure of the AFE supercell. We found that although the AFE displacement is large, the mass enhancement is still negligible ($\beta_e$≈1.1 and $\beta_h$≈1), in other words, the electronic response to such displacements (deformation potential) must be small.

This weak response of band-edge states to Ti off-center displacement can be understood by considering the symmetry mismatch between the orbitals making up the VBM and CBM, and the symmetry of the Ti displacement mode: In the cubic primitive-cell of BaTiO$_3$ (non-electric single-f.u. model without displacements or tilting), the system has the space group of $Pm\bar{3}m$, where, according to the molecular orbital theory for octahedral O$_h$ symmetry [131], the CBM is pure Ti-$d$ orbital (irreducible representation T$_{2g}$), and the VBM is made up by O-2$p$ + Ti-4$p$ orbitals (irreducible representation T$_{1u}$). Considering that the Ti-4$p$ orbital is very high in energy, the VBM is almost pure O-$p$ (T$_{1u}$). On the other hand, our supercell model of the AFE state has a large $M_2^-$ distortion mode due to *Ti displacements* (amplitude = 0.15 Å, irreducible representation B$_{1u}$ for D$_{4h}$ symmetry) whereas *tilting and rotation* amplitudes are all smaller than 6×10$^{-4}$ Å. Since the VBM and CBM are not B$_{1u}$ symmetric, they do not respond to the Ti displacements, thus the AFE supercell only shows negligible mass enhancement for these states.

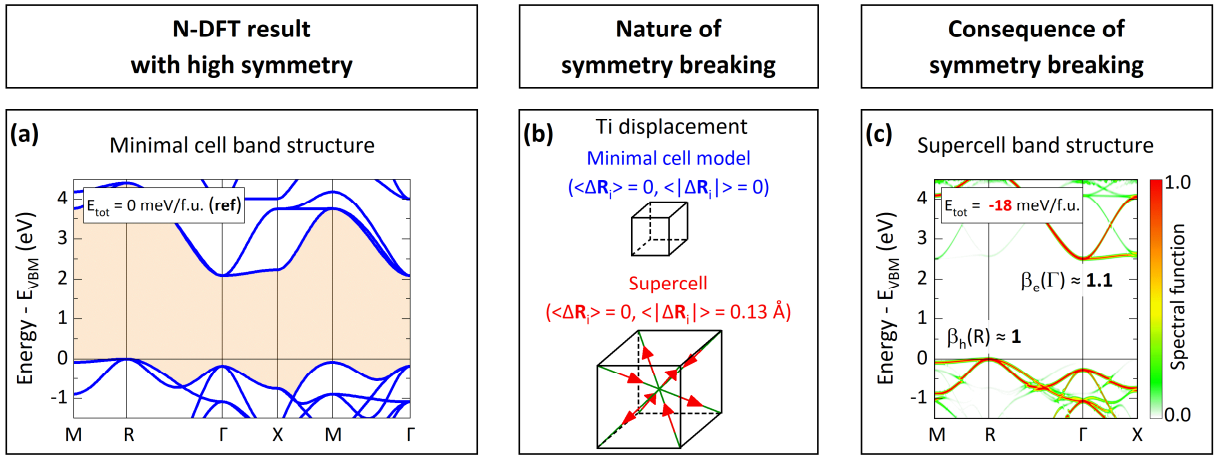

**Figure 11 (2 columns) | Mass enhancements in AFE cubic phase of BaTiO₃.** (a) shows the band structure from the same N-DFT restriction as in previous literature [127–129], namely a single-cell, no Ti-atom displacement model, using SCAN functional. (b) gives the distribution of Ti-atom displacements symmetry breaking: Upper part in (b) shows that the minimal-cell model BaTiO$_3$ does not have any Ti-atom displacement (Δ**R**=0 for each Ti atom), while the lower part in (b) shows that the AFE supercell (8 f.u.) has a unique displacement pattern, where the 8 Ti atoms move along 8 <111> directions. In the AFE supercell, the net polarization is zero (<Δ**R**>=0), but the local polarization on each Ti site is large (<|Δ**R**|>=0.13 Å). (c) shows the unfolded band structure of the 8-f.u. AFE supercell BaTiO$_3$ with the same SCAN method. Masses in (c) are calculated via the second derivative of *E* vs ***k***, which gives $\beta_e$≈1.1 and $\beta_h$≈1.

The PE phase has been modeled by a 32-f.u. supercell. After the atomic relaxation, all Ti atoms have developed non-zero, non-uniformed local polarizations, as shown by the red curved in Figure 12(b). It can be seen that the Ti displacement forms a distribution not only on amplitude but also on directions. We note that the net displacement (vector summation of all Ti displacements) is not zero (0.1 Å per f.u.), because we do not force any net-polarization condition during the supercell relaxation, and the supercell is, in fact, weak FE. Nevertheless, the nature of a distribution of static (non-thermal), different displacement in the supercell is by itself different than the low-T FE phase. Solving the band structure and doing band unfolding (Figure 12(c)) shows small mass enhancement of $\beta_e$≈1.1 and $\beta_h$≈1. The results of both AFE and PE seem consistent in that the coupling of polarization to VBM and CBM states is weak, illustrating cases that the ferroelectric displacement has a negligible effect on mass enhancement.

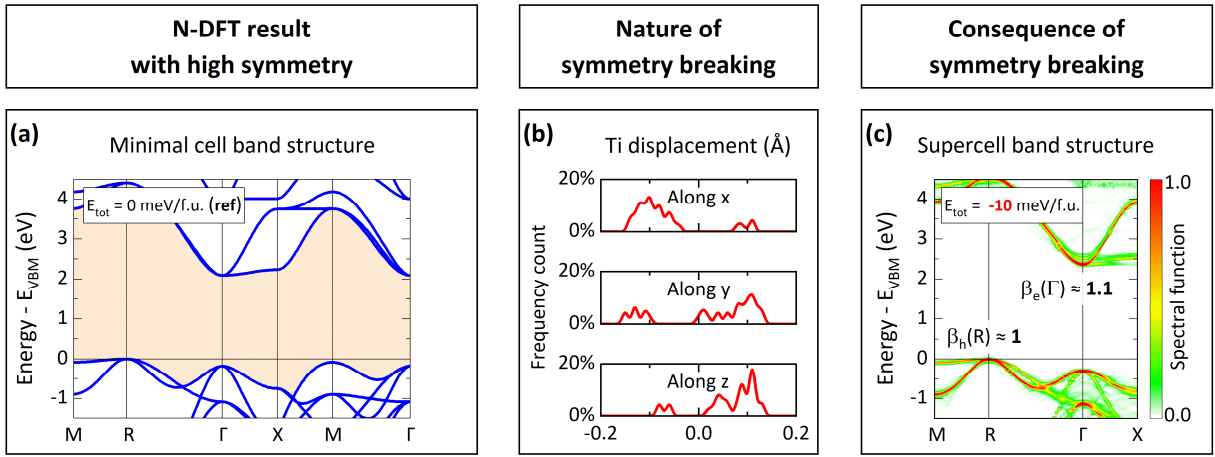

**Figure 12 (2 columns) | Mass enhancements in the PE cubic phase of BaTiO₃.** (a) shows the band structure from the same N-DFT restriction as in previous literature [127–129], identical to Figure 11(a). (b) gives the distribution of Ti-atom displacements in the 32-f.u. PE supercell after atomic relaxation along x, y, and z directions ([100], [010], and [001] directions). (c) shows the unfolded band structure of the 32-f.u. PE supercell BaTiO₃ with the same SCAN method. Masses in (c) are calculated via the second derivative of $E$ vs $k$, which gives $\beta_e \approx 1.1$ and $\beta_h \approx 1$.

## VI. Conclusions

Symmetry-breaking DFT captures many of the mass enhancement effects previously attributed exclusively to strong electronic correlations under restricted symmetry. The current approach provides an intuitive explanation for the alternative physical origins of mass enhancement, as summarized in Fig. 1, including (1) spin symmetry-breaking effect and (2) positional symmetry-breaking effects. Such symmetry-broken or distortion effects are common in both *d*- and *p*-electron perovskites. There are cases where the coupling of distortions to the electronic states at band edges are weak, causing negligible mass renormalization (*e.g.*, SrTiO₃ where the distortion is small, or BaTiO₃ where only Ti displacement not tilting lowers the total energy), yet other cases where the distortions and their coupling are strong, leading to large enhancement factors (*e.g.*, SrVO₃, CsPbI₃, LaMnO₃, and SrBiO₃), even by the single-determinant mean-field DFT method, which sometimes are even comparable to the mass enhancement suggested by the high-order dynamical electron-electron correlation theory. The presence of mass enhancement in the considered systems is not necessarily evidence for the exclusive need for strong dynamically correlated methodologies [15–24].

## Acknowledgement

The work at the University of Colorado at Boulder was supported by the U.S. Department of Energy, Office of Science, Basic Energy Sciences, Materials Sciences and Engineering Division, under Grant No. DE-SC0010467 to the University of Colorado. The *ab initio* calculations in this work were performed using resources of the National Energy Research Scientific Computing Center, which is supported by the Office of Science of the U.S. Department of Energy.

## Appendix A: *DFT details.*

To calculate the total energy, band structure, and effective mass, the plane wave pseudopotential DFT method, as implemented in the VASP software package [132,133] has been used. (i) For transition metal oxides with localized orbitals $SrTiO_3$, $BaTiO_3$, and $LaMnO_3$, the SCAN functional [134], used previously [97,135] has been applied. (ii) For metallic $SrVO_3$, such SCAN functional could not reach a self-converged charge density in our large PM cubic supercell (640 atom/f.u.); therefore, for all $SrVO_3$ cells, the Perdew–Burke-Ernzerhof (PBE) functional + $U$ (with $U$=1.25 eV on V-d orbitals) used previously [27] has been applied instead. Figure 13 shows the difference of DOS between the SCAN and PBE+U DFT results for a smaller 320-atom supercell of PM cubic phase $SrVO_3$. It can be seen from Figure 13 that the PBE+U and SCAN functionals give (1) very similar DOS at Fermi level, and (2) very similar bandwidths; they, therefore, should predict very similar mass enhancement factors. (iii) The *s-p* bonded halide $CsPbI_3$ has been calculated using the PBE functional. Note that no spin-orbit coupling (SOC) has been considered for $CsPbI_3$ in this work because the PBE functional will give too small gap (~0.1 eV) if implied together with SOC; a better agreement with experimentally observed gap can be achieved by using SOC with a hybrid functional such as HSE, which is however beyond the scope of this work. (iv) The *s-p* bonded oxide $SrBiO_3$ has been calculated using the PBE with spin-orbit coupling effect.

For each compound, we have applied the DFT lattice constants obtained from the minimal cell model to all supercell calculations: a=3.83 Å (cubic $SrVO_3$), a=6.27 Å (cubic $CsPbI_3$), a=5.51 Å b=5.81 Å and c=7.64 Å ($LaMnO_3$ with Jahn-Teller-like distortion), a=5.57 Å and c=7.87 Å ($LaMnO_3$ without Jahn-Teller distortion), a=6.01 Å b=6.20 Å and c=10.49 Å (monoclinic $SrBiO_3$), a=4.52 Å (cubic $SrBiO_3$), a=3.91 Å (cubic $SrTiO_3$), and a=4.03 Å (cubic $BaTiO_3$). To minimize the numerical error, for all cells of the same compound in the same phase (e.g., $SrVO_3$ single-f.u. cubic primitive cell vs. $SrVO_3$ 64-f.u. cubic supercell), their total energies are calculated using an uniformed energy cutoff for the plane-wave basis set, an uniformed tolerance for total energy convergence ($10^{-8}$ eV/atom), and an *equivalent set of k-point* sampling in BZ for every cell (equivalent to a 12×12×12 Γ-centered k-mesh in the primitive BZ).

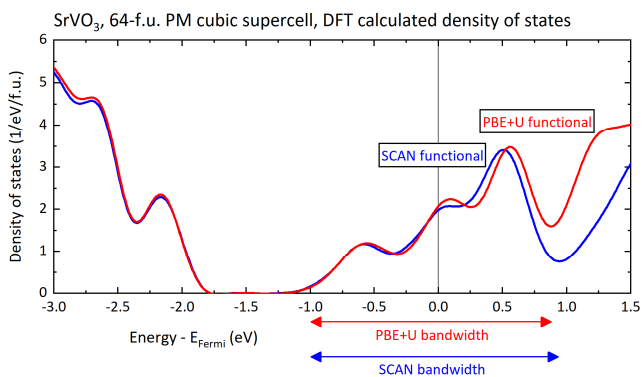

**Figure 13 (1 column) |** Comparison of DOS from DFT calculations using PBE+U (red curve) and SCAN (blue curve) functionals for the 64-f.u. (320-atom) PM cubic supercell of SrVO$_3$. The two functionals show remarkably similar DOS at Fermi level and very similar bandwidths (denoted by the red and blue arrows).

## Appendix B: *Magnetic structure of SrVO$_3$.*

Table II summarizes the DFT total energies of different magnetic orders of cubic SrVO$_3$: NM – nonmagnetic, FM – ferromagnetic, AFM-G – G-type antiferromagnetic, AFM-A – A-type antiferromagnetic, and PM – collinear paramagnetic phase from SQS method. All magnetic structures show significant energy lowering than the nonmagnetic model, but very similar total energy with each other. The results agree with the experimental observation that SrVO$_3$ shows no magnetic order down to T=0 K.

**Table II |** Summary of DFT total energy for different collinear magnetic structures of cubic SrVO$_3$. The same PBE+U method has been applied for all SrVO$_3$ calculations. All structures have the same lattice constant and atomic positions and only differ in the way how up and down spin moments occupy vanadium sites. Magnetic orders NM, FM, AFM (G-type and A-type), and PM denote nonmagnetic, ferromagnetic, antiferromagnetic and paramagnetic (from SQS method), respectively. The DFT total energy of the NM phase has been chosen as the reference (0 eV).

| Cubic SrVO$_3$ magnetic phase | Total energy (meV/f.u.) |
|---|---|
| **NM** (1 f.u.) | 0 (reference) |
| **FM** (1 f.u.) | -15 |
| **AFM-G** (8 f.u.) | -31 |
| **AFM-A** (2 f.u.) | -26 |
| **PM** (64 f.u.) | -30 |
| **PM** (128 f.u.) | -31 |